\documentclass[aps, prab, reprint, showpacs, amsmath, amssymb]{revtex4-1}

\usepackage{amsmath,amscd}
\usepackage{bm}
\usepackage{color,soul}
\usepackage[english]{babel}
\usepackage{graphicx}

\renewcommand{\Re}{\mathop{\mathrm{Re}}\nolimits}
\renewcommand{\Im}{\mathop{\mathrm{Im}}\nolimits}

\begin{document}
\title{ Radiation of a Charge Exiting Open-Ended Waveguide with Dielectric Filling }

\author{Sergey N. Galyamin}
\email{s.galyamin@spbu.ru}
\affiliation{Saint Petersburg State University, 7/9 Universitetskaya nab., St. Petersburg, 199034 Russia}

\author{Andrey V. Tyukhtin}
\email{a.tyuhtin@spbu.ru}
\affiliation{Saint Petersburg State University, 7/9 Universitetskaya nab., St. Petersburg, 199034 Russia}

\author{Viktor V. Vorobev}
\affiliation{Saint Petersburg State University, 7/9 Universitetskaya nab., St. Petersburg, 199034 Russia}

\author{Alexandra A. Grigoreva}
\affiliation{Saint Petersburg State University, 7/9 Universitetskaya nab., St. Petersburg, 199034 Russia}

\author{Alexander S. Aryshev}
\affiliation{KEK: High Energy Accelerator Research Organization, 1-1 Oho, Tsukuba, Ibaraki, 305-0801 Japan}

\date{\today}

\begin{abstract}
We consider a semi-infinite open-ended cylindrical waveguide with uniform dielectric filling placed into collinear infinite vacuum waveguide with larger radius. 
Electromagnetic field produced by a point charge or Gaussian bunch moving along structure's axis from the dielectric waveguide into the vacuum one is investigated. 
We utilize the modified residue-calculus technique and obtain rigorous analytical solution of the problem by determining coefficients of mode excitation in each subarea of the structure. 
Numerical simulations in CST Particle Studio are also performed and an excellent agreement between analytical and simulated results is shown.
The main attention is paid to analysis of Cherenkov radiation generated in the inner dielectric waveguide and penetrated into vacuum regions of the outer waveguide.
The discussed structure can be used for generation of Terahertz radiation by modulated bunches (bunch trains)
by means of high-order Cherenkov modes.
In this case, numerical simulations becomes difficult while the developed analytical technique allows for efficient calculation of the radiation characteristics.   

\end{abstract}


\maketitle

\section{ Introduction \label{sec:intro} }

In recent years, an essential interest is observed in the area of contemporary sources of Terahertz (THz) radiation based on beam-driven waveguide structures loaded with dielectric.
Despite of the fact that both ordinary vacuum THz devices (such as classical backward wave oscillator) are widely available and other mechanisms for THz sources are discussed (see, e.g., Refs.~%
\cite{Wil06,Wen13,Kaz17}%
), beam-driven sources are still extremely attractive due to extraordinary THz radiation peak power~%
\cite{OShea16}. 
According to this idea, Cherenkov radiation should be generated by well-controlled electron bunch passed through a waveguide structure with dielectric filling and open aperture~%
\cite{Ant13,Ant15}.
The electron bunch should be modulated so that a high-order Cherenkov frequency is excited, therefore allowing the use of, for example, mm-sized waveguides for THz generation.   
Another challenge here is efficient extraction of the radiation from inside the structure into free space.
The possibilities for using the non-orthogonal end cut for this purpose were theoretically estimated~%
\cite{GTAB14}
and experimentally confirmed~%
\cite{Ant16}.
Nevertheless, rigorous solution for the electromagnetic (EM) field produced by a charged particle bunch passing from the open-ended circular waveguide with dielectric filling is still missing even in the case of orthogonal end cut.
In particular, such a solution is required for determination of the area of applicability of the approximate technique used in~%
\cite{GTAB14}
and it's possible improvement.

General theory for analysis of radiation from open-ended waveguide structures was actively developed during several preceding decades~%
\cite{Vainb, Mittrab, BGalst00}. 
Typically, the theory of EM processes for the discussed waveguide discontinuity (open end) was constructed for vacuum case and excitation by single waveguide mode, however, vacuum structures excited by a moving charge were also partially investigated~%
\cite{KH86, KPV87, Pal90, T14, IGTT14}.
It should be also noted here that analytical approaches becomes essentially more complicated while they deal with the structures containing dielectric inclusions~%
\cite{Mittrab, VBlarMittra69, VZh78}.   

In a series of recent papers, we started rigorous investigation of the aforementioned problem on EM field in a circular open-ended waveguide with orthogonal cut and dielectric filling excited by the field of a moving charged particle bunch~%
\cite{GTABproc16, GVAGTA17Proc, GTV17, GTVGA18, GTVA18}.
In these publications, the semi-infinite waveguide was placed (embedded) into collinear infinite vacuum waveguide with larger radius.
Therefore the considered structure is further referred to as ``embedded'' structure.
Based on a very good agreement between analytical and simulated (using both COMSOL and CST) results observed in the aforementioned papers one can conclude that convenient analytical technique for solution of the described problem was fully approved.

It should be noted that the closed geometry has several advantages compared to the opened one for theory, simulations and possible experiments.
First, closed structure possesses discrete mode spectrum, thus simplifying analytical consideration.
Second, finite area of EM field existence allows efficient simulations without the need of extremely large amount of computational resources.
Third, real experiments on THz generation from the open end can be conducted in circular vacuum chamber, the latter is described by the outer waveguide in the theoretical model.     


However, Cherenkov radiation which is responsible for the aforementioned high-power THz emission and therefore is of most interest in the structure under consideration was not described in details. 
In the present paper, we give the detailed analytical solution for EM field generated by a charged particle bunch in the ``embedded'' structure loaded with dielectric.
We apply the modified residue-calculus technique~%
residue-calculus technique~%
\cite{VBlarMittra69}
and describe penetration of Cherenkov radiation into vacuum regions of the structure.

The paper is organized as follows.
After the Introduction (Sec.~%
\ref{sec:intro}%
), we present rigorous solution of the problem (Sec.~%
\ref{sec:anal}%
).
Note that this section contains only final analytical results while details of calculations and intermediate derivations are placed into three appendices (App.~%
\ref{App1},
\ref{App2}
and
\ref{App3}%
) succeeding the main text.
Section~%
\ref{sec:num}
presents numerical results visualizing the obtained rigorous formulas, simulated results (via CST PS package)
and comparison between them.
The Conclusion (Sec.~%
\ref{sec:concl}
finishes the paper.  
   
%
%
\begin{figure}[t]
\centering
\includegraphics[width=8.5cm]{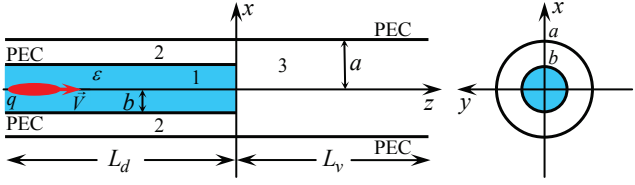}
\caption{\label{fig:geom} Geometry of the problem and main notations.
$ L_v $ 
and 
$ L_d $ 
are lengths of vacuum and dielectric parts of the model, correspondingly. 
They are infinite in theory and finite in simulations.}
\end{figure}
%
%

\section{ Analytical results \label{sec:anal} }

Geometry of the problem under consideration is shown in Fig.~%
\ref{fig:geom}. 
A semi-infinite perfectly conducting circular waveguide with radius 
$ b $ 
filled with a homogeneous dielectric 
($\varepsilon > 1$) 
is put into a concentric infinite waveguide with radius 
$ a > b $. 
The structure is excited by a point charge 
$ q $
moving along 
$ z $%
-axis with constant velocity 
$ \vec{ V } = V \vec{ e }_{ z } = \beta c \vec{ e }_{ z } $
(%
$ c $
is the light speed in vacuum
).
Corresponding charge density 
$ \rho $
and current density
$ \vec{j} = j \vec{e}_z $
have the form
\begin{align}
\label{eq:roj}
\rho = q \delta( x ) \delta( y ) \delta( z - V t ),
\quad
j = V \rho.
\end{align}
Unless otherwise specified, analytical results presented below correspond to the source~%
\eqref{eq:roj}.
These results can be easily generalized for the case of a bunch being infinitesimally thin in 
$ xy $%
-plane, similar to~%
\eqref{eq:roj},
but having arbitrary charge distribution
$ \eta( z - V t ) $
along 
$ z $
(longitudinal) direction.
In this case, charge and current densities,
$ \rho_b $
and
$ \vec{ j }_b = j_b \vec{ e }_z$
are:
\begin{align}
\label{eq:bunch}
\rho_b = q \delta( x ) \delta( y ) \eta( z - V t ),
\quad
j_b = V \rho_b. 
\end{align}

As can be easily shown, to obtain formulas related to the case of the bunch~%
\eqref{eq:bunch}
one should substitute
\begin{equation}
\label{eq:subst}
q \to 2 \pi q \tilde{ \eta }( \omega / V ),
\end{equation}
where
$ \tilde{ \eta }( \omega / V ) $
is the Fourier transform
\begin{equation}
\label{eq:tildeeta}
\tilde{ \eta }( \xi ) =
( 2 \pi )^{-1}
\int\nolimits_{-\infty}^{+\infty}
\eta(\zeta)
e^{ -i \xi \zeta }
\, d \zeta.
\end{equation}
calculated for
$ \xi = \omega / V $.
For example, in the case of Gaussian bunch with the rms half-length
$ \sigma $,
\begin{equation}
\label{eq:gaussian}
\eta_{ \mathrm{ G } }( z - V t )
=
\frac{ 1 }{ \sqrt{ 2 \pi } \sigma }
\exp{ \left( \frac{ - ( z - V t )^2 }{ 2 \sigma^2 } \right) },
\end{equation}
and one should substitute
\begin{equation}
\label{eq:substg}
q
\to
q \exp{ \left( - \frac{ \omega^2 }{ \omega_{ \sigma }^2 } \right) },
\quad
\omega_{ \sigma } = \frac{ \sqrt{ 2 } V }{ \sigma }.
\end{equation}
In this case, the largest essential frequency in the spectrum 
$ \omega_{ \mathrm{ max } } $
is determined so that Gaussian exponential term in
\eqref{eq:substg}
results in certain predetermined attenuation for
$ \omega = \omega_{ \mathrm{ max } } $.
Typical attenuation (for example, used in CST~PS code by default) is 
$ -20 $%
dB, that is
\begin{equation}
\label{eq:dB}
20\mathrm{lg}
\left[
1 
\left/
\exp{ \left( - \left. \omega_{ \mathrm{ max } }^2 \right/ \omega_{ \sigma }^2  \right) }
\right.
\right]
=
-20.
\end{equation} 
This result in 
$ \omega_{ \mathrm{ max } }  = \omega_{\sigma} \sqrt{ \ln 10 } \approx 1.5 \omega_{\sigma} $.

Further the cylindrical frame 
$ r $,
$ \phi $,
$ z $
(associated with the Cartesian frame shown in Fig.~%
\ref{fig:geom}%
) is used.
The problem will be solved in the frequency domain, so that Fourier harmonic 
$ H_{ \omega \phi } $ 
will be determined. 
Other nonzero field components are calculated as follows:
\begin{align}
\label{eq:Hphi2Er} 
&E_{ \omega r } = c( i \omega \varepsilon )^{-1}
{\partial H_{\omega \phi } / \partial z},
\\
\label{eq:Hphi2Ez} 
&E_{\omega z} = -c ( i \omega \varepsilon r )^{ - 1 }
\left[
H_{ \omega \phi }
+
r { \partial H_{ \omega \phi } / \partial r }
\right].
\end{align}
Time-domain field dependencies are calculated using the inverse Fourier transform formulas which can be transformed to the following form~%
\cite{TG08,GT10}%
:
\begin{equation}
\label{eq:Hphitime}
H_{ \phi } ( r, z, t ) 
= 
2{ \Re } \int\nolimits_{ 0 }^{ +\infty } H_{ \omega \phi } e^{ -i \omega t } \, d \omega. 
\end{equation}
On the basis of Eq.~%
\eqref{eq:Hphitime}, 
it is sufficient to consider only positive frequencies in the spectrum.

\subsection{Incident field \label{sec:analinc}}

Fourier harmonic of the magnetic component of the incident field has the following form~%
\cite{B62}: 

\begin{equation} 
\label{eq:inc}
H_{ \omega \phi }^{ (i) } =
\begin{cases} 
H_{ \omega \phi }^{ (i1) }, &\text{ for $ z < 0 $, } \\
H_{ \omega \phi }^{ (i3) }, &\text{ for $ z > 0 $. }
\end{cases}
\end{equation}
Here
\begin{equation} 
\label{eq:inc1}
H_{ \omega \phi }^{ ( i1 ) }
{=}
\frac{ i q { s } }{ 2c } 
\left[ 
H_{ 1 }^{ ( 1 ) } 
( r { s } ) 
{-} 
\frac{ H_{ 0 }^{ ( 1 ) } ( b s ) }{ J_{ 0 } ( b s ) } 
J_{ 1 } ( r s )
\right]
e^{ \frac{ i \omega z }{ V } }, 
\end{equation}
$ s(\omega) = \sqrt{ \omega^{ 2 } V^{ -2 } ( \varepsilon \beta^{ 2 } -1 ) } $, 
$ \Im{ s } > 0 $,
\begin{equation} 
\label{eq:inc3}
H_{ \omega \phi }^{ ( i3 ) }
{ = }
\frac{ i q s_0 }{ 2c } 
\left[ 
H_{ 1 }^{ ( 1 ) }( r s_0 ) 
{-} 
\frac{ H_{ 0 }^{ ( 1 ) } ( a s_0 ) }{ J_{ 0 } ( a s_0 ) } 
J_{ 1 } ( r s_0 )
\right]
e^{ \frac{ i \omega z }{ V } }, 
\end{equation}
$ s_{ 0 }( \omega ) = \sqrt{ \omega^{ 2 } V^{ - 2 } ( \beta^{ 2 } -1 ) } $, 
$ \Im{ s_0 } > 0 $, 
$ J_{ 0,1 } $
are Bessel functions,
$ H_{ 0,1 }^{ ( 1 ) } $
are Hankel functions of the first order. 
Equation~%
\eqref{eq:inc1}
represents the total field of a point charge~%
\eqref{eq:roj}
uniformly moving in regular waveguide of radius 
$ b $ 
filled with dielectric
$ \varepsilon $.
Equation~%
\eqref{eq:inc3}
represents the total field of the same charge moving in regular vacuum waveguide with radius 
$ a $.


For
$ \varepsilon \beta ^ 2 > 1 $,
incident field in the inner dielectric waveguide
$ H_{ \omega \phi }^{ ( i1 ) } $
contains field of Cherenkov radiation (so called ``wakefield'').
Wakefield is the part of
\eqref{eq:inc1}
with the discrete frequency spectrum, namely the finite set of real ``Cherenkov frequencies''
$ \omega_{ l }^{ \mathrm{ Ch } } $
which correspond to real positive poles of the expression~%
\eqref{eq:inc1}. 
These poles are determined by the following equation: 
\begin{equation}
\label{eq:Chfreq}
J_{0} ( b s )
=
0
\quad
\Rightarrow
\quad
s\!\left( \omega_{ l }^{ \mathrm{ Ch } } \right)
=
j_{ 0l } / b,
\end{equation}
where
$ j_{ 0 l } $
is the zero of the zero-order Bessel function,
$ J_0( j_{ 0 l } ) { = } 0 $, 
$ l = 1, 2, \ldots $.
It can be shown (for example, by the limiting process from the case with dissipation taken into account in dielectric) that the integration path in~%
\eqref{eq:Hphitime}
passes real ``Cherenkov poles'' from above.  
Therefore, these poles contribute to the incident field only behind the charge, i.e. for
$ \zeta  = z - Vt < 0 $.
Contributions of these poles (residues) can be calculated:
\begin{equation}
\label{eq:incVCR}
H_{ \phi }^{ \mathrm{ Ch } ( i1 ) }( r, z, t )
=
\sum\nolimits_{ l = 1 }^{ \infty }
H_{ \phi l }^{ \mathrm{ Ch } ( i1 ) }( r, z, t ),
\end{equation}
where
\begin{equation}
\label{eq:incVCRm}
\begin{aligned}
&H_{ \phi l }^{ \mathrm{ Ch } ( i1 ) }( r, z, t )
{ = }
2\Re
\left[
( { - } 2 \pi i )
\mathrm{ Res }_{ \omega_{ l }^{ \mathrm{ Ch } } }
H_{ \omega \phi }^{ ( i1 ) }
e^{ -i \omega_{ l }^{ \mathrm{ Ch } } t }
\right]
{ = } \\
&{ = }
2\Re
\left[
\frac{ \pi q j_{ 0 l } }{ b c } 
\frac{ H_{ 0 }^{ ( 1 ) } ( j_{ 0 l } ) }{ J_{ 1 } ( j_{ 0 l } ) } 
J_{ 1 } ( r j_{ 0 l } / b )
e^{ i \omega_{ l }^{ \mathrm{ Ch } } \zeta / V }
\right].
\end{aligned} 
\end{equation}
%
In the case of Gaussian bunch~%
\eqref{eq:gaussian}
due to vanishing exponential term~%
\eqref{eq:substg}
in the spectrum, high-order Cherenkov frequencies are strongly suppressed, therefore one can obtain monochromatic Cherenkov radiation for long enough bunch (this is also true for arbitrary finite length bunch).

In the geometry under consideration (see Fig.~%
\ref{fig:geom}%
), the waveguide with dielectric filling has an open end, therefore Cherenkov radiation generated inside the inner waveguide will penetrate both coaxial part of the structure (area 2 in Fig.~%
\ref{fig:geom}%
)
and wide vacuum part (area 3 in in Fig.~%
\ref{fig:geom}%
).
Note that penetration of Cherenkov radiation through simple plane infinite interface between two media accompanying generation of transition radiation was investigated previously~%
\cite{Gar58,GTKS09,GT10,GT11}. 
In the case under consideration, the process of penetration occures due to the diffraction mechanism.
This process is of main interest in this paper and it can be described by the theory presented below.  
%
%
\subsection{ Scattered field \label{sec:analscatt} }

The unknown additional (scattered) field propagating from the boundary in the domains 1, 2 and 3 can be presented as standard series over corresponding waveguide modes~%
\cite{Mittrab}%
:
\begin{equation}
\label{eq:Hphiinner}
H_{\omega \phi }^{ ( 1 ) } ( r, z ) = \sum\limits_{ m = 1 }^{ \infty } 
B_{ m } J_{ 1 } ( r j_{ 0 m } / b ) e^{ \kappa_{ z m }^{ ( 1 ) } z },
\end{equation}
\begin{equation}
\label{eq:Hphiwide}
H_{ \omega \phi }^{ ( 3 ) } ( r, z ) = \sum\limits_{ m = 1 }^{ \infty } A_{ m } J_{ 1 } ( r j_{ 0 m } / a )
e^{ - \gamma_{ z m }^{ ( 3 ) } z },
\end{equation}
\begin{equation}
\label{eq:Hphicoax}
H_{ \omega \phi }^{ ( 2 ) } ( r, z ) =
C_{ 0 } r^{ - 1 } e^{ \gamma_{ z 0 }^{ ( 2 ) } z } 
+
\sum\limits_{ m = 1 }^{ \infty } C_{ m } Z_{ m } ( r \chi _{ m } ) e^{ \gamma_{ z m }^{ ( 2 ) } z }.
\end{equation}
Note that the first term in the right-hand side of 
\eqref{eq:Hphicoax}
represents the TEM wave with
$ E_{ \omega z } = 0 $, 
in accordance with~%
\eqref{eq:Hphi2Ez}.
Here
\begin{equation}
\label{eq:Z}
Z_{ m }( \xi ) = J_{ 1 } ( \xi ) - N_{ 1 } ( \xi ) J_{ 0 } ( a \chi_{ m } ) N_{ 0 }^{ -1 } ( a \chi_{ m } )
\end{equation}
is the transversal eigenfunction of the coaxial region (area 2 in Fig.~%
\ref{fig:geom}),
$ \chi_{ m } > 0 $
is the solution of the dispersion relation for the area 2,
\begin{equation}
\label{eq:coaxdisp}
J_{ 0 } ( b \chi_{ p } ) N_{ 0 }( a \chi_{ p } ) - J_{ 0 } ( a \chi_{ p } ) N_{ 0 }( b \chi_{ p } ) = 0,
\end{equation}
$ N_{ 0 } $
is the Neumann function.
Propagation constants are:
\begin{equation}
\label{eq:kappa1}
\kappa_{ z m }^{ ( 1 ) } = \sqrt{ j_{ 0 m }^{ 2 } b^{ - 2 } - \varepsilon k_{ 0 }^{ 2 } },
\end{equation}
\begin{equation}
\label{eq:gamma3}
\gamma_{ z m }^{ ( 3 ) } = \sqrt{ j_{ 0 m }^{ 2 } a^{ - 2 } - k_{ 0 }^{ 2 } },
\end{equation}
\begin{equation}
\label{eq:gamma2}
\gamma_{ z 0 }^{ ( 2 ) } = -i k_0,
\quad
\gamma_{ z m }^{ ( 2 ) } = \sqrt{ \chi_{ m }^2 - k_{ 0 }^{ 2 } },
\end{equation}
where
$ k_{0} = \omega / c $,
$ \Re{ \kappa_{ z m }^{ ( 1 ) } } > 0 $,
$ \Re{ \gamma_{ z m }^{ ( 2, 3 ) } } > 0 $,
$ m = 1, 2, \ldots $.

For readers' convenience, below we discuss the way for solving the problem under consideration just briefly and present the main resulting formulas only.
Rather cumbersome details of calculations needed for deep understanding of the used technique are moved into the Appendices.
Corresponding references are given in the text.  

Performing matching of the components 
$ H_{ \omega \phi } $ 
and 
$ E_{\omega r} $ 
for 
$ z = 0 $, 
and eliminating the 
$ r $%
-dependence from the resulting relations, 
after certain analytical transformations we obtain infinite systems for unknown coefficients 
$ \{ A_m \} $,
$ \{ B_m \} $
and
$ \{ C_n \} $
($ n = 0, 1, 2, \ldots $)
of mode decompositions
\eqref{eq:Hphiwide},
\eqref{eq:Hphiinner}
and
\eqref{eq:Hphicoax},
correspondingly.
This procedure is explained in detail in Appendix~%
\ref{App1}%
, where obtained systems
\eqref{eq:sys1},
\eqref{eq:sys2},
\eqref{eq:sys3}
and 
\eqref{eq:sys4}
are presented.
Using the modified residue-calculus technique~%
\cite{Mittrab,GTV17}, 
these systems can be solved by constructing specific complex-valued function
$ f( w ) $.
This procedure is described in detail in the Appendix~%
\ref{App2}.
Finally, the coefficients can be expressed through
$ f( w ) $
as follows:
\begin{equation}
\label{eq:A}
A_{ m } = \frac{ {\rm Res}_{ \gamma_{ zm }^{ ( 3 ) } } f( w ) }{ J_0( b j_{ 0m } / a  ) j_{ 0m } / a },   
\end{equation}
\begin{equation}
\label{eq:B}
\begin{aligned}
B_{ m } &= 
\frac{ \varepsilon \gamma_{ zm }^{ ( 1 ) } + \kappa_{ zm }^{ ( 1 ) } } 
{ 2 b J_1 ( j_{ 0m } ) \gamma_{ zm }^{ ( 1 ) } \kappa_{ zm }^{ ( 1 ) } }
\left[ \frac{ i q }{ 2 c b } \left( R_m F_{ dm }^{+} + F_{ dm }^{ - } \right) - \right.  \\
&\left. - R_m F_{ vm }^{-} - F_{ vm }^{ + } - R_m f( \gamma_{ zm }^{ ( 1 ) } ) - f( -\gamma _{ zm }^{ ( 1 ) } ) 
\vphantom{ \frac{ i q }{ 2 c b } }
\right],
\end{aligned}   
\end{equation}
\begin{equation}
\label{eq:C0}
C_{ 0 } = \frac{ f( -\gamma _{ z 0 }^{ ( 2 ) } ) }{ 2 \gamma_{ z 0 }^{ ( 2 ) } { \rm ln }( a / b ) },
\end{equation}
\begin{equation}
\label{eq:C}
C_{ m } = \frac{ f( -\gamma_{ z m }^{ ( 2 ) } ) }
{ 2 \gamma_{ zm }^{ ( 2 ) }
\left[
\frac{ a^2 Z_m^2( a \chi_m  )}{ 2 b Z_m( b \chi_m  ) }
- \frac{ b }{ 2 } Z_m( b \chi_m )
\right] }.
\end{equation}
Here
\begin{equation}
\label{eq:gamma1}
\gamma_{ zm }^{ ( 1 ) } = \sqrt{ j_{ 0m }^{ 2 } b^{ -2 } - k_{ 0 }^{ 2 } },
\quad
\Re{ \gamma_{ zm }^{ ( 1 ) } } > 0
\end{equation}
is the propagation constant of the area 1 in the case of 
$ \varepsilon = 1 $,
\begin{equation}
\label{eq:R}
R_m = \frac{ \varepsilon \gamma_{ zm }^{ ( 1 ) } - \kappa_{ zm }^{ ( 1 ) } }{ \varepsilon \gamma_{ zm }^{ ( 1 ) } + \kappa_{ zm }^{ ( 1 ) } },
\end{equation}
\begin{equation}
\label{eq:Fd}
F_{ dm }^{ \pm } 
= 
\frac{ 2i j_{ 0 p } }{ \pi b }
\frac{\frac{ \omega }{ iV \varepsilon } \pm \gamma_{zm}^{(1)} }{ s^2 - ( j_{0m} / b )^2 },
\end{equation}
\begin{equation}
\label{eq:Fv}
F_{ vm }^{ \pm } 
= 
\frac{ 2i j_{ 0p } ( \pi b )^{ - 1 } } { \frac{ \omega }{ iV } \pm \gamma_{ zm }^{ ( 1 ) } }.
\end{equation}
Function
$ f( w ) $
is determined as follows:
\begin{equation}
\label{eq:f}
f( w ) = \frac{ P g( w ) }{ w - \frac{ \omega }{ i V } },
\end{equation} 
\begin{equation}
\label{eq:g}
g( w )
{ = }
\frac{ ( w { - } \gamma_{ z 0 }^{ ( 2 ) } )
\!\!
\prod\limits_{ n { = } 1 }^{ \infty }
\!\!
\left( 1 {-} \frac{ w }{ \gamma_{ z n }^{ ( 2 ) } } \right) }
{ \prod\nolimits_{ m { = } 1 }^{ \infty } \left( 1 { - } \frac{ w }{ \gamma_{ z m }^{ ( 3 ) } } \right) } 
\prod_{ s { = } 1 }^{ \infty }
\left( 1 { - } \frac{ w }{ \Gamma_{ s } } 
\right)
Q( w ),
\end{equation} 
\begin{equation}
\label{eq:Q} 
Q( w ) 
{ = }
\exp \left[
{ - }
\frac{ w }{ \pi } 
\left( b \ln \left( \frac{ b }{ a { - } b } \right) 
{ + } 
a \ln \left( \frac{ a { - } b }{ a } \right) \right) \right],
\end{equation} 
\begin{equation}
\label{eq:P}
P = \frac{ i q }{ 2 c } \frac{ i s_0^{ 2 } h_{ 0 } }{ g( \omega / ( i V ) ) },
\end{equation}
\begin{equation}
\label{eq:h0}
h_{ 0 } = N_{ 0 } ( b s_{ 0 } ) - N_{ 0 }( a s_{ 0 } ) J_{ 0 } ( b s_{ 0 } ) / J_{ 0 } ( a s_{ 0 } ).
\end{equation}
%

The correct construction of the function 
$ f( w ) $
\eqref{eq:f}
is the key point of the residue-calculus technique.
The details of this procedure is described in detail in the Appendix~%
\ref{App2}.
In particular, one should determine zeros
$ \{ \Gamma_m \} $
shifted with respect to zeros of vacuum problem
$ \{ \gamma_{ zm }^{ ( 1 ) } \} $:
\begin{equation}
\label{eq:Gamma}
\Gamma_m
=
\gamma_{ z m }^{ ( 1 ) } + \frac{ \pi }{ b } \Delta_m,
\end{equation}
where the set
$ \{ \Delta_m \} $
determines the unknown shifts.
Shifted zeros of 
$ f( w ) $
are the distinguishing feature of the problem with dielectric and this fact complicates significantly the solution (compared to the vacuum case) because the set
$ \{ \Gamma_m \} $
is determined for each distinct frequency
$ \omega $.
Since this is connected with iterative solution of a certain complicated nonlinear system (see Eq.~%
\eqref{eq:infsyst}
in Appendix~%
\ref{App2}%
), it is difficult to obtain the field spectrum
$ H_{ \omega \phi } $
for significant range of frequencies
$ [ 0, \omega_{\max} ] $.
Therefore, it is difficult to calculate full time dependencies for the field components using inverse Fourier transform formulas~%
\eqref{eq:Hphitime}.

However, the detailed analysis of the function 
$ f( w ) $
shows that it contains the same Cherenkov poles
$ \left\{ \omega_{ l }^{ \mathrm{ Ch } } \right\} $
as the incident field~%
\eqref{eq:inc1}
does (see the Appendix~%
\ref{App3}%
).
Therefore, Cherenkov radiation penetrated all vacuum areas of the structure which is described by contribution of these poles (residues) can be easily calculated, similar to Eq.~%
\eqref{eq:incVCR}.
For example, Cherenkov radiation penetrated areas 2 and 3 (which is of most interest) can be expressed as follows:
\begin{equation} 
\label{eq:CTR}
H_{ \phi }^{ \mathrm{Ch} ( \alpha ) }( r, z, t )
=
\sum\nolimits_{ l = 1 }^{\infty}
H_{ \phi l }^{ \mathrm{Ch} ( \alpha ) }( r, z, t ),
\end{equation}
\begin{equation}
\label{eq:CTRm}
H_{ \phi l }^{ \mathrm{Ch} ( \alpha ) }(r, z, t)
=
2\Re
\left[
-2 \pi i
\mathrm{Res}_{ \omega_{ l }^{ \mathrm{ Ch } } }
H_{ \omega \phi }^{ ( \alpha ) }
e^{ -i \omega_{ l }^{ \mathrm{ Ch } } t }
\right],
\end{equation}
$ \alpha = 2,3 $
(means corresponding subarea of the structure).
Note that each summand in Eq.~%
\eqref{eq:CTRm}
depends on Fourier transform of the scattered field
$ H_{ \omega \phi }^{ ( \alpha ) } $
which is presented as infinite series over waveguide modes
(Eqs.~%
\eqref{eq:Hphiwide}
and
\eqref{eq:Hphicoax}%
).
Since we suppose that~%
\eqref{eq:CTRm}
describes radiation, these series should be truncated to contain only propagating modes for given Cherenkov frequency.

Let us discuss the procedure to calculate the contribution of 
$ \omega_{ l }^{ \mathrm{ Ch } } $
for given 
$ l $.
Since 
$ \{ A_m \} $
and
$ \{ C_n \} $
have the pole for
$ \omega = \omega_{ l }^{ \mathrm{ Ch } } $,
then
\begin{equation}
\label{eq:CTRpole}
H_{ \omega \phi }^{ (\alpha) }
\approx
\frac{ \mathrm{Res}_{ \omega_{ l }^{ \mathrm{ Ch } } }
H_{ \omega \phi }^{ ( \alpha ) } }
{ \omega - \omega_{ l }^{ \mathrm{ Ch } } },
\quad
\omega \to \omega_{ l }^{ \mathrm{ Ch } }.
\end{equation}
Here, the term in the numerator is the residue to be found.
If we suppose that dielectric possesses some small dissipation, 
$ \varepsilon = \varepsilon^{ \prime } + i \varepsilon^{ \prime\prime } $,
then Cherenkov pole also becomes complex:
\begin{equation}
\label{eq:Chfreqcomplex}
\omega_{ l }^{ \mathrm{ Ch } }
=
\omega_{ l }^{ \mathrm{ Ch } \prime }
+
i \omega_{ l }^{ \mathrm{ Ch } \prime\prime },
\quad
\omega_{ l }^{ \mathrm{ Ch } \prime\prime } < 0.
\end{equation}    
Using numerical procedure described in Appendix~%
\ref{App2},
we calculate
$ H_{ \omega \phi }^{ ( \alpha ) } $
for
$ \omega = \omega_{ l }^{ \mathrm{ Ch } \prime } $,
therefore
\begin{equation}
\label{eq:CTRpolecalc}
\mathrm{Res}_{ \omega_{ l }^{ \mathrm{ Ch } } }
H_{ \omega \phi }^{ ( \alpha ) }
=
-i \omega_{ l }^{ \mathrm{ Ch } \prime\prime }
\left. H_{ \omega \phi }^{ ( \alpha ) } \right|_{ \omega = \omega_{ l }^{ \mathrm{ Ch } \prime } }
\end{equation}
In this way one can calculate contributions of all the essential Cherenkov poles relatively simple and fast.
Corresponding examples are represented below in Sec.~%
\ref{sec:num}.

\begin{figure}[b]
\centering
\includegraphics[width=7.5cm]{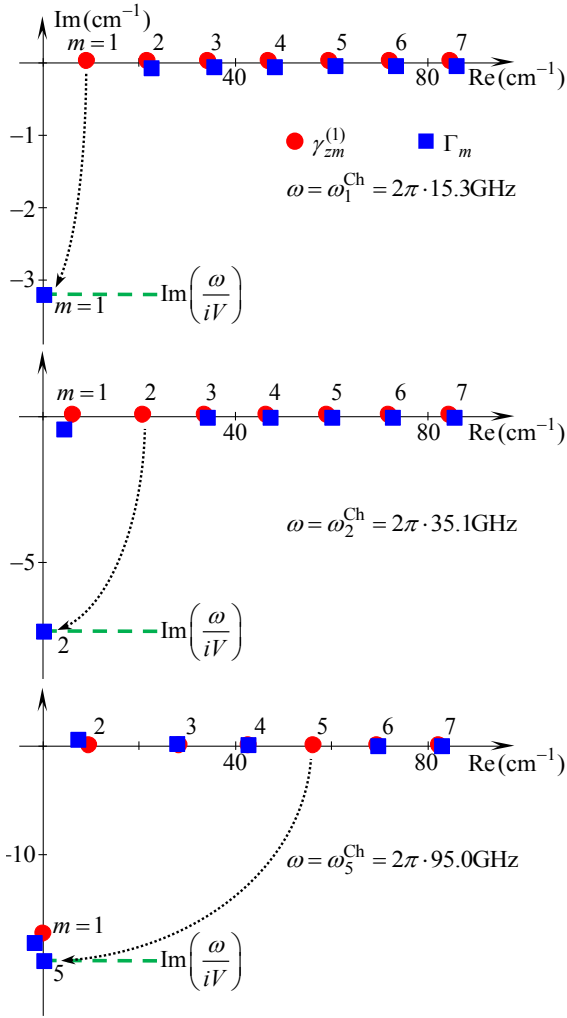}
\caption{\label{fig:shiz}
Comparison between shifted and ``unshifted'' zeros 
$ \gamma_{ z m }^{ ( 1 ) }~( \mathrm{ cm }^{ -1 } ) $
and 
$ \Gamma_m~( \mathrm{ cm }^{ -1 } ) $,  
$ m = 1,2, \ldots 7 $ 
for three frequencies of Cherenkov radiation
$ \omega = \omega_{l}^{ \mathrm{Ch} \prime } $,
$ l = 1, 2, 5 $.
Structure parameters are:
$ b = 0.25${\rm cm}, 
$ a = 0.9 ${\rm cm}, 
$ \varepsilon = 10 + i 10^{ -5 } $.
}
\end{figure}
%

\section{ Numerical Results and Discussion \label{sec:num} }

Here we present numerical results obtained via rigorous formulas of the previous section and results of direct numerical simulation in CST PS~{\textregistered} package (with the use of wakefield solver).
For simulation, we constructed the model with finite 
$ L_v $
and 
$ L_d $
(see Fig.~%
\ref{fig:geom}%
)
and open boundary conditions for 
$ z = -L_d $,
$ z = L_v $.
Also small finite thickness of the inner waveguide wall 
$ d_w \ll a, b $
was taken into account in simulations, i.e. we supposed that coaxial area 2 is determined by inequality 
$ b + d_w < r < a $,
$ z < 0 $.
The adaptive meshing procedure was utilized to obtain optimal simulation parameters and stable results, this point will be explained below using representative example (see Fig.~%
\ref{fig:freqtend_eps10_25_50}%
).

%
\begin{table}[t]
\centering
\caption{Comparison between shifted and ``unshifted'' zeros (%
$ \left. \gamma_{ zm }^{ ( 1 ) } \right/ \Gamma_m $%
) for 
$ m = 1, 2, \ldots 7 $
and three Cherenkov frequencies
$ \omega_{ l }^{ \mathrm{ Ch } } $%
,
$ l = 1, 2, 5 $.  
}
\label{tab:shifted} 
\begin{tabular}{ l | c | c | c }
\hline
$ m $  & $\omega_{1}^{ \mathrm{Ch} }$  &  $\omega_{2}^{ \mathrm{Ch} }$      & $\omega_{5}^{ \mathrm{Ch} }$ \\
\hline
1      & 9.07/-3.21$i$       & 6.19/4.31-0.45$i$   & -17.43$i$/-2.01-18.18$i$ \\ \hline
2      & 21.85/22.53-0.08$i$ & 20.82/-7.36$i$      & 9.55/7.11+0.54$i$ \\ \hline
3      & 34.47/35.38-0.08$i$ & 33.82/34.17-0.05$i$ & 28.32/27.76+0.16$i$ \\ \hline
4      & 47.06/48.10-0.07$i$ & 46.59/47.13-0.06$i$ & 42.76/42.55+0.06$i$ \\ \hline
5      & 59.64/60.77-0.06$i$ & 59.27/59.95-0.06$i$ & 56.31/-19.91$i$\\ \hline
6      & 72.21/73.41-0.05$i$ & 71.91/72.69-0.05$i$ & 69.49/69.63-0.04$i$ \\ \hline
7      & 84.79/86.03-0.05$i$ & 84.53/85.38-0.05$i$ & 82.48/82.73-0.06$i$ \\ \hline
\end{tabular}
\end{table}
%

First, we clarify the statement concerning the frequency spectrum of the fields in vacuum areas of the structure.
Figure~%
\ref{fig:shiz}
shows position of the first seven shifted zeros
$ \Gamma_m $
and 
``unshifted'' zeros
$ \gamma_{ zm }^{ ( 1 ) } $
on the complex plane calculated for three Cherenkov frequencies
\eqref{eq:Chfreq}.
These results are supplemented by Table~%
\ref{tab:shifted}
where corresponding numerical values (%
$ \gamma_{ zm }^{ ( 1 ) } / \Gamma_m $%
) are presented with the 
$ 0.01 $
relative accuracy.
Note that since a small dissipation in dielectric is taken into account (%
$ \varepsilon^{ \prime\prime } / \varepsilon^{ \prime } = 10^{ -6 } $%
), Cherenkov frequencies are complex values, as Eq.~%
\eqref{eq:Chfreqcomplex}
indicates.
The discussed calculations are performed for
$ \omega = \omega_{ l }^{ \mathrm{ Ch } \prime } $.

%
\begin{figure}[b]
\centering
\includegraphics[width=0.9\linewidth]{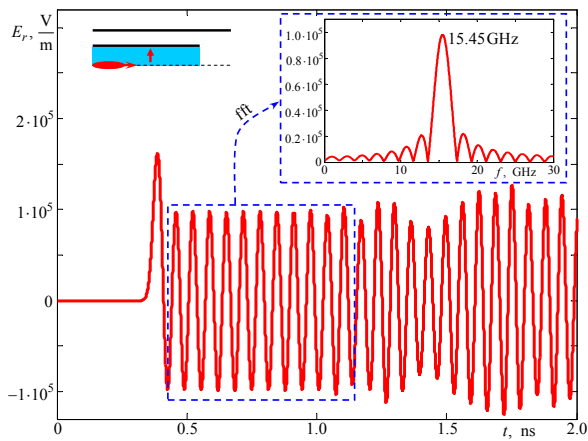}
\caption{\label{fig:probeinner_eps10_25_50}%
Behavior of the electric field 
$ E_r $ 
over time (CST result) on the probe in the inner waveguide: 
$ z = -1 $~cm, 
$ r = 0.125 $~cm.
Time 
$ t = 0 $ 
corresponds to the moment when bunch center is in the plane 
$ z = 0 $.
Structure parameters: 
$ b = 0.25 $~cm, 
$ a = 0.5 $~cm, 
$ \varepsilon = 10 $, 
$ L_d = 35 $~cm, 
$ L_v = 50 $~cm, 
inner waveguide wall thickness is 
$ 0.001 $~cm. 
Gaussian bunch parameters: 
$ q = 1 $~nC, 
$ \beta = 0.9999 $, 
$ \sigma = 0.5 $~cm. 
}
\end{figure}
%

As one can see from Fig.~%
\ref{fig:shiz}%
, the majority of presented 
$ \Gamma_m $ 
are weakly shifted with respect to
$ \gamma_{ z m }^{ ( 1 ) } $
excluding 
$ \Gamma_l $
(with the number of Cherenkov frequency under consideration).
This
$ \Gamma_l $
is shifted dramatically so that it becomes purely imaginary while initial
$ \gamma_{ z l }^{ ( 1 ) } $
was purely real.
Moreover, one can learn from Table~%
\ref{tab:shifted}
that the equality
\begin{equation}
\label{eq:polesequal}
\Gamma_l( \omega_{ l }^{ \mathrm{ Ch } } )
\approx
\left.
\omega_{ l }^{ \mathrm{ Ch } }
\right/
( i V )
\end{equation}
is fulfilled with high accuracy, because
\begin{equation}
\label{eq:Chpolesnum}
\frac{ \omega_{ 1 }^{ \mathrm{ Ch } } }{ i V } 
\approx
-3.21i,
\;
\frac{ \omega_{ 2 }^{ \mathrm{ Ch } } }{ i V } 
\approx
-7.36i,
\;
\frac{ \omega_{ 5 }^{ \mathrm{ Ch } } }{ i V } 
\approx
-19.91i.
\end{equation}
Note that in Eq.~%
\eqref{eq:Chpolesnum}
we present numerical values with 
$ 0.01 $
relative accuracy, similar to Table~%
\ref{tab:shifted}.

As it is shown in Appendix~%
\ref{App2},
Eq.~%
\eqref{eq:Chpolesnum}
leads to conclusion that all sets of unknown coefficients possess poles for Cherenkov frequencies. 
Contribution of these poles in vacuum areas of the structure can be calculated using Eq.~%
\eqref{eq:CTRpolecalc}. 

%
\begin{figure}[t]
\centering
\includegraphics[width=0.9\linewidth]{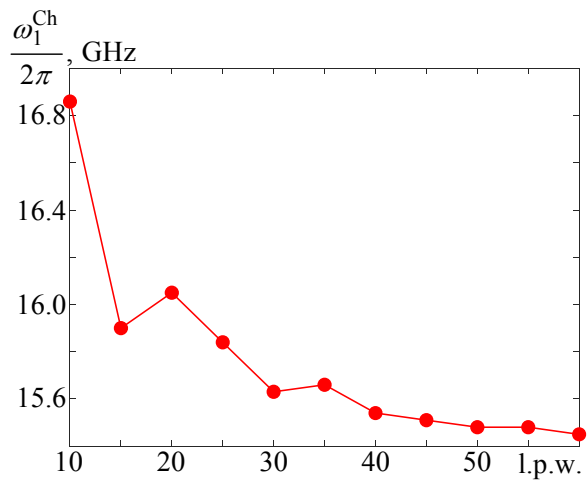}
\caption{\label{fig:freqtend_eps10_25_50}%
Dependence of the simulated first Cherenkov frequency  
$ \omega_{1}^{ \mathrm{Ch} } / ( 2 \pi ) $
(GHz)
on number of lines per wavelength (l.p.w.).
Problem parameters are the same as in Fig.~%
\ref{fig:probeinner_eps10_25_50}%
. 
}
\end{figure}
%

\subsection{ Single bunch field }

Here we present numerical results illustrating the field behavior in different subareas of the structure.
For all figures, radius of the inner waveguide is the same,
$ b = 0.25 $%
~cm.
The structure is excited by single relativistic Gaussian bunch 
\eqref{eq:gaussian}.

\begin{figure}[b]
\centering
\includegraphics[width=0.9\linewidth]{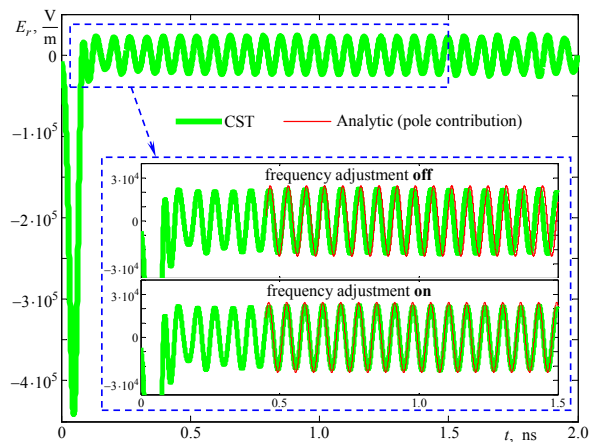}
\caption{\label{fig:probecoax_eps10_25_50} Behavior of the electric field (%
$ E_r $%
) over time on the probe in the coaxial area: 
$ z = -1 $~cm, 
$ r = 0.35 $~cm.
Structure and bunch parameters are the same as in Fig.~%
\ref{fig:probeinner_eps10_25_50}
}
\end{figure}

\begin{figure}[t]
\centering
\includegraphics[width=0.9\linewidth]{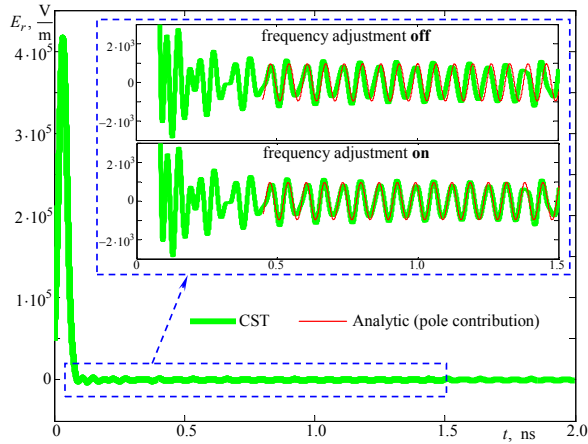}
\caption{\label{fig:probewide_eps10_25_50}%
Behavior of the electric field (%
$ E_r $%
) over time on the probe in the wide vacuum waveguide: 
$ z = 1 $~cm, 
$ r = 0.35 $~cm.
Structure and bunch parameters are the same as in Fig.~%
\ref{fig:probeinner_eps10_25_50}
}
\end{figure}

\begin{figure}[b]
\centering
\includegraphics[width=8cm]{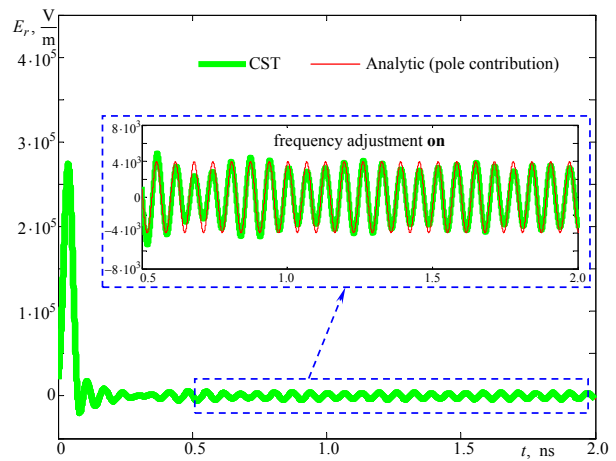}
\caption{\label{fig:probewide_eps10_25_90}%
Behavior of the electric field (%
$ E_r $%
) over time on the probe in the wide vacuum waveguide: 
$ z = 1 $~cm, 
$ r = 0.35 $~cm.
Outer waveguide radius
$ a = 0.9 $%
~cm, other parameters are the same as in Fig.~%
\ref{fig:probeinner_eps10_25_50}
}
\end{figure}

\begin{figure}[t]
\centering
\includegraphics[width=0.9\linewidth]{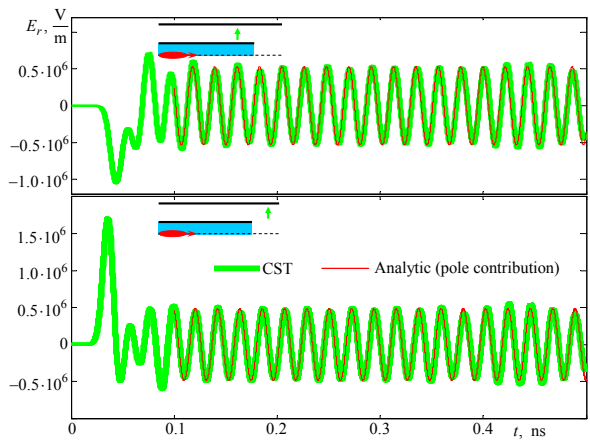}
\caption{\label{fig:probecoaxwide_eps2_25_50}%
Behavior of the electric field
$ E_r $
over time on the symmetrical probes in areas 2 and 3: 
$ z = \pm 1 $~cm, 
$ r = 0.35 $~cm.
Permittivity 
$ \varepsilon = 2 $, 
other parameters are the same as in Fig.~%
\ref{fig:probeinner_eps10_25_50}
}
\end{figure}

For Figs.~%
\ref{fig:probeinner_eps10_25_50}
--
\ref{fig:probecoaxwide_eps2_25_50}%
, the bunch length 
$ \sigma $ 
is chosen so that only the first Cherenkov frequency
$ \omega_{1}^{ \mathrm{ Ch } } $
lies within essential part of frequency spectrum 
$ [ 0, \, \omega_{ \mathrm{ max } } ] $%
~%
\eqref{eq:dB} 
while higher Cherenkov frequencies are suppressed by attenuating Gaussian term
\eqref{eq:substg}.
In this case we expect monochromatic both Cherenkov radiation in the inner waveguide and Cherenkov radiation penetrating areas 2 and 3.

%
%

Figure~%
\ref{fig:probeinner_eps10_25_50}~%
shows transverse electric field 
$ E_r $
from the probe located in the inner waveguide for the case of relatively large permittivity,
$ \varepsilon = 10 $,
and
$ a = 0.5 $~cm.
The part of the signal enclosed in the dashed line rectangle (%
$ 0.4~\mathrm{ns} < t < 1.1~\mathrm{ ns } $%
) should be interpreted as the field of Cherenkov radiation.
The Fourier spectrum of this part of the signal shown in the inset of Fig.~%
\ref{fig:probeinner_eps10_25_50}
has a strong peak for frequency
$ 15.45 $~%
GHz, this is Cherenkov radiation frequency obtained in the numerical experiment.
For shortness, this result for simulated frequency will be referred to as ``experimental'' result throughout this section. 

It should be noted that mentioned peak is used for adaptive meshing procedure in CST simulation: mesh is refined (number of lines per wavelength is increased) until the position of the peak becomes stable, i.e. relative difference in position is less than 
$ 0.002 $
for two consequent passes.
Figure~%
\ref{fig:freqtend_eps10_25_50}
illustrates this procedure.
It shows typical dependence of the experimental Cherenkov frequency on the number of meshlines per wavelength (this is standard parameter defining mesh density in CST; wavelength is understood as the minimal wavelength which corresponds to
$ \omega_{ \mathrm{ max } }$%
).
As one can see, for rare mesh the experimental frequency is considerably larger compared to the theoretical value (%
$ 15.31 $~%
GHz).
For
$ 60 $
lines per wavelength the relative difference is sufficiently small therefore procedure is stopped, the obtained frequency differs from the theoretical by less then one percent.
Comparison between first Cherenkov frequencies, theoretical and experimental, for all structures discussed below is shown in Table~%
\ref{tab:chfreq}.
Theoretical value of Cherenkov frequency does not depend on radius of the outer waveguide
$ a $,
but this is not the case for simulations due to the change in mesh with change in 
$ a $.
In all considered cases, relative difference between theoretical and experimental values is around 1 percent.
As one can see below, this small difference matters in comparison of the field behavior.

%
\begin{table}[t]
\centering
\caption{Comparison between analytical and experimental (CST) first Cherenkov frequency
$ \omega_{1}^{ \mathrm{Ch} } $
(Analytical / Experimental) for 
$ b = 0.25 $~cm.  
}
\label{tab:chfreq} 
\begin{tabular}{ c | c | c }
\hline
               & $ \varepsilon = 10 $  &  $ \varepsilon = 2 $ \\
\hline
$ a = 0.5$~cm   & 15.31~GHz / 15.45~GHz   &  45.9~GHz / 46.13~GHz   \\ \hline
$ a = 0.9$~cm   & 15.31~GHz / 15.48~GHz   &  45.9~GHz / 46.27~GHz   \\ \hline
\end{tabular}
\end{table}
%

Figure~%
\ref{fig:probecoax_eps10_25_50}
shows CST simulated signal from the probe located in area 2 of the structure with
$ \varepsilon = 10 $
and
$ a = 0.5 $~cm.
Solid (green) line corresponds to the field obtained via simulation in CST~PS~{\textregistered} code.
According to the CST curve, with an increase in time
$ t $, 
a strong peak corresponding to the ``image'' of the bunch can be seen first (this effect was discussed in details in the case of similar vacuum structure in Ref.~%
\cite{GTVA18}%
). 
After that, some transition process connected with diffraction radiation occurs.
For large enough time (%
$ t \gtrsim 0.5 $%
ns) we see the stationary harmonic process.
Top inset in Fig.~%
\ref{fig:probecoax_eps10_25_50}
shows magnified part of the CST curve compared with theoretical curve corresponding to contribution of the first Cherenkov pole,
i.e. summand with
$ l = 1 $
and
$ \alpha = 2 $
in
\eqref{eq:CTR}
(red line).
Magnitudes correlate well but due to the difference in frequency the curves diverge for large enough time.
If we manually adjust the frequency in analytical formulas, i.e. substitute the analytical Cherenkov frequency with the simulated one (see Table~%
\ref{tab:chfreq}%
),
we will obtain an excellent coincidence between the curves shown in the bottom inset in Fig.~%
\ref{fig:probecoax_eps10_25_50}.
The described frequency substitution is further called ``frequency adjustment''.
Based on presented comparison and the tendency for experimental Cherenkov frequency (see Fig.~%
\ref{fig:freqtend_eps10_25_50}%
) one can conclude on both correctness of the used analytical approach and stable operation of the simulation code for fine enough mesh.
  

%
%
%
%
Figure~%
\ref{fig:probewide_eps10_25_50}~%
shows similar comparison (for the same structure) but for probe located in the area 3 (wide vacuum waveguide).
Again, after the frequency adjustment applied the curves correlate very well.
Further for all figures the frequency adjustment will be used by default.
Note that for given 
$ a $
even the first mode in area 3 is evanescent therefore magnitude of Cherenkov radiation is extremely small. 
Figure~%
\ref{fig:probewide_eps10_25_90},~%
illustrates the case of 
$ \varepsilon = 10 $
and
larger radius of the outer waveguide,
$ a = 0.9 $~cm.
In this case, Cherenkov radiation penetrated area 3 is more expressed and again it is described very well by analytical formulas. 

Figure~%
\ref{fig:probecoaxwide_eps2_25_50},~%
shows signals from symmetrical probes in coaxial and vacuum waveguide areas for the structure with lower permittivity (%
$ \varepsilon = 2 $%
) and correspondingly higher Cherenkov frequency.
%
%
Again, one can see that pole contribution calculated theoretically describes Cherenkov radiation penetrated vacuum parts of the structure very well.

\subsection{ Bunch train field }

Here we illustrate the possibilities of the described approach for calculation of Cherenkov radiation at high-order modes.
According to the idea of beam-driven THz source described in Sec.~%
\ref{sec:intro}%
, THz frequencies can be generated in mm-sized waveguides by charged particle bunches with proper charge modulation, i.e. by bunch trains~%
\cite{Ant15}.
If we denote by 
$ \tilde{ \eta }_{ G } $
the Fourier spectrum of a single Gaussian bunch 
$ \eta_{ \mathrm{ G } } $~%
\eqref{eq:gaussian}%
, we obtain from~%
\eqref{eq:tildeeta}:
\begin{equation}
\label{eq:gaussian_spectrum}
\tilde{ \eta }_{ \mathrm{ G } }( \xi ) 
=
( 2 \pi )^{ -1 } \exp{ \left( -\xi^2 \sigma^2 / 2 \right) }.
\end{equation}
The sequence of 
$ 2 M + 1 $
identical Gaussian bunches spaced by 
$ L $
and carrying the same total charge has the following longitudinal charge distribution:
\begin{equation}
\label{eq:gaussian_sequence}
\eta_{ \mathrm{ Seq } }( z - V t )
=
\frac{ 1 }{ 2 M + 1 }
\sum\limits_{ m = -M }^{ M }
\eta_{ \mathrm{ G } }( z - V t + m L ),
\end{equation}
and the following spectrum, in accordance with~%
\eqref{eq:tildeeta}:
\begin{equation}
\label{eq:gaussian_sequence_spectrum}
\begin{aligned}
\tilde{ \eta }_{ \mathrm{ Seq } }( \xi )
&=
\frac{ \tilde{ \eta }_{ \mathrm{ G } }( \xi ) }
{ 2 M + 1 }
\times \\
& \times
\left[
1
+
2 \cos{ \left(  \frac{ \xi L ( M + 1 ) }{ 2 } \right) }
\frac{ \sin{ \left( \xi L M / 2 \right) } }{ \sin{ \left( \xi L / 2 \right) } }
\right].
\end{aligned}
\end{equation}

\begin{figure}[t]
\centering
\includegraphics[width=0.9\linewidth]{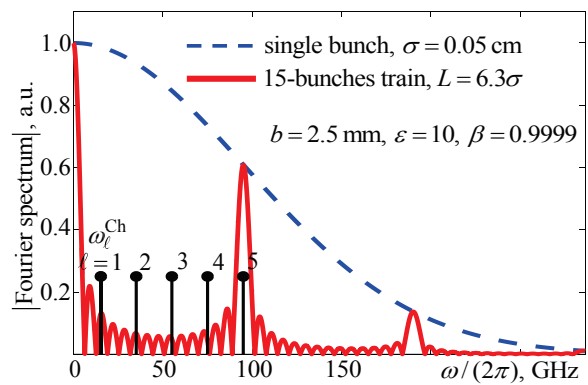}
\caption{\label{fig:15bunch_train} 
Fourier spectrum of a single Gaussian bunch and that of a 15 bunches train with 
$ L > 2 \sigma $ 
spacing.
Black markers show Cherenkov frequencies
$ \omega_{ l }^{ \mathrm{ Ch } } $.
Bunches parameters:
$ q = 1 $~nC, 
$ \beta = 0.9999 $,
$ \sigma = 0.05 $~cm,
$ L = 6.3 \sigma $.
Cherenkov frequencies are calculated for the inner waveguide with
$ b = 0.25 $~cm
filled with dielectric with
$ \varepsilon = 10 $.
}
\end{figure}

Figure~%
\ref{fig:15bunch_train} 
shows comparison of a single Gaussian bunch Fourier spectrum~%
\eqref{eq:gaussian_spectrum}
with the spectrum of a bunch train~%
\eqref{eq:gaussian_sequence_spectrum}
of 15 (%
$ M = 7 $%
) identical bunches with spacing 
$ L > 2 \sigma $.
Both functions are calculated for
$ \xi = \omega / V $.
Cherenkov frequencies
\eqref{eq:Chfreq}
for a mm-sized waveguide are also shown. 
Parameters 
$ \sigma $
and
$ L $
are chosen so that the bunch train spectrum has the expressed maximum exactly at Cherenkov frequency 
$ \omega^{ \mathrm{ Ch } }_{ 5 } $.
Therefore, this bunch train excites effectively the 5-th Cherenkov mode with the frequency around 0.1~THz falling in the lower part of THz range. 
In the same manner, other Cherenkov frequencies can be generated separately. 

\begin{figure*}[t]
\centering
\includegraphics[width=0.8\linewidth]{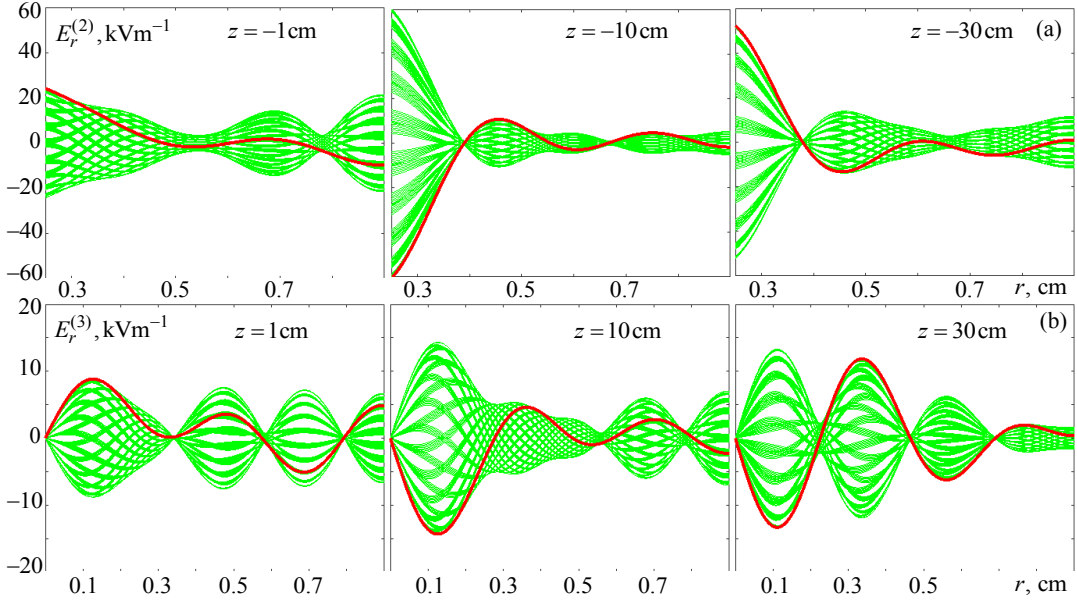}
\caption{\label{fig:15bunch_train_field} 
Cherenkov radiation field (%
$ E_r $
component) at 5-th Cherenkov frequency,
$ \omega^{ \mathrm{ Ch } }_{ 5 } \approx 2 \pi { \cdot } 95$%
GHz, in vacuum regions of the structure: in the coaxial area (a) and in wide vacuum waveguide (b).
Each thin (green) curve shows the 
$ E_r^{ \mathrm{ Ch } } $
as a function of
$ r $
at given time moment
$ t $ 
and given
$ z $.
In total, each plot contains 
$ 151 $
curves covering the 
$ 1.5 $%
~ns time range with 
$ 0.01 $%
~ns interval.
Solid (red) curve corrresponds to the global field maximum over the cross-section.
Parameters of the structure and the bunch train are the same as in Fig.~%
\ref{fig:15bunch_train}
so that the 5-th Cherenkov frequency is effectively generated.
}
\end{figure*}

It should be noted that simulation of EM field produced by such bunch trains is rather complicated in CST PS package.
In particular, according to Fig.~%
\ref{fig:freqtend_eps10_25_50}%
, number of meshlines per wavelength required for adequate convergence should be increased considerably.
Another issue here is manual determination of bunch profile corresponding to the discussed bunch train.   
On the contrary, the presented analytical technique allows computation of the EM field properties relatively simple and fast which is illustrated below by Fig.~%
\ref{fig:15bunch_train_field}.
%
%

%
%

Figure~%
\ref{fig:15bunch_train_field}
shows behaviour of 
$ E_r $
component of Cherenkov radiation at 5-th Cherenkov frequency,
$ \omega^{ \mathrm{ Ch } }_{ 5 } \approx 2 \pi { \cdot } 95$%
GHz, in vacuum regions of the structure.
Recall that this radiation is generated in the inner waveguide and penetrated vacuum sections of the structure by means of diffraction mechanism.
Figure~%
\ref{fig:15bunch_train_field}%
(a) shows
$ E_r $ 
field in three cross-sections of the coaxial region (area 2 in Fig.~%
\ref{fig:geom}%
) while Fig.~%
\ref{fig:15bunch_train_field}%
(b) shows 
$ E_r $ field in three cross-sections of the wide vacuum waveguide (area 3 in Fig.~%
\ref{fig:geom}%
).
Each thin (green) curve shows the 
$ E_r^{ \mathrm{ Ch } } $
as a function of
$ r $
at a given time moment
$ t $ 
and given
$ z $.
In total, each plot contains 
$ 151 $
curves covering the 
$ 1.5 $%
ns time range with 
$ 0.01 $%
ns interval.
The highlighted solid (red) curve is the curve which provides the maximum field magnitude over the cross-section.
Since at the given Cherenkov frequency both coaxial waveguide and wide vacuum waveguide supports 5 propagating modes, field behaviour is rather complicated.
As one can see, maximum field in coaxial region is always on the inner waveguide wall.
In the wide waveguide, global field maximum is typically at the first or second local maximum.

\section{ Conclusion\label{sec:concl} }
We have considered radiation produced by single Gaussian bunch exiting the open end of a cylindrical waveguide with uniform dielectric filling in the case where this waveguide is put into concentric vacuum infinite waveguide of larger radius.
Based on residue-calculus technique, we have constructed the rigorous theory of the electromagnetic process in this structure.
Based on this theory, Cherenkov radiation exiting from dielectric waveguide into vacuum parts of the structure, which is of essential interest in the context of beam driven radiation sources, can be calculated easily and fast.
We also have performed numerical simulation of the process in CST~PS code.
It has been shown that simulated Cherenkov radiation spectral peak has correct frequency for only dense enough mesh.
In our simulations, we have reached mesh density around 60 lines per minimal wavelength in the spectrum, resulting in around
$ 1 \% $ 
difference between theoretical and numerical frequencies.
In this case, numerical and analytical results for Cherenkov radiation coincided very well therefore proving both the correctness of rigorous approach and good convergence of numerical procedure.

Moreover, we have considered generation of high-order Cherenkov modes by modulated bunches (bunch trains) in vacuum regions of the structure.
Since trains of short bunches generate relatevely high frequencies, correct numerical simulations will require large amount of calculating resources. 
In this case, the presented rigorous approach allowing convenient analysis of the EM field across the structure will be the preferred method of investigation.
As an example, we have calculated spatiotemporal distribution of Cherenkov field at the 5-th Cherenkov frequency (around 
$0.1$%
THz) generated in vacuum regions of mm-sized embedded structure with dielectric filling of the inner waveguide.


%

\appendix

\section{\label{App1} Infinite systems for mode decomposition coefficients
$\{ A_m \}$, $\{ B_m \}$ and $\{ C_n \}$}

Boundary conditions in the plane
$ z = 0 $
result in the following relations:
\begin{align}
\label{eq:Hphicont1}
H_{\omega \phi }^{(1)} ( r, 0 ) &=
H_{\omega \phi }^{(3)} ( r, 0 )
& \text{ for $ 0 \le r \le b $ }, 
\\
\label{eq:Hphicont2}
H_{\omega \phi }^{(2)} ( r, 0 ) &=
H_{\omega \phi }^{(3)} ( r, 0 )
& \text{ for $ b \le r \le a $ },
\end{align}
\begin{align}
\label{eq:Erocont1}
\frac{ \partial H_{\omega \phi }^{(1)} ( r, z ) }{ \varepsilon \,\, \partial z } &=
\left. \frac{ \partial H_{\omega \phi }^{(3)} ( r, z ) }{ \partial z } \right|_{ z = 0 }
&\text{ for }
&0 \le r \le b, 
\\
\label{eq:Erocont2}
\frac{ \partial H_{\omega \phi }^{(2)} ( r, z ) }{ \partial z } &=
\left. \frac{ \partial H_{\omega \phi }^{(3)} ( r, z ) }{ \partial z } \right|_{ z = 0}
& \text{ for }
&b \le r \le a.
\end{align}
To eliminate dependence on
$ r $,
we substitute 
\eqref{eq:Hphiinner}
and
\eqref{eq:Hphiwide}
into
\eqref{eq:Hphicont1}
and
\eqref{eq:Erocont1},
integrate obtained relations over 
$ 0 < r < b $ 
with the weight function
$ r J_1( r j_{0p} / b ) $,
$ p = 1, 2, \ldots $
and utilize the following properties
\cite{pbm1}:
\begin{equation}
\label{eq:int1}
\int\limits_{0}^{b}
r J_1 \left( \frac{ r j_{0p} }{ b } \right) J_1 \left( \frac{ r j_{0m} }{ b } \right ) dr
=
\frac{ b^2 J_1^2( j_{0p} ) \delta_{pm} }{ 2 },
\end{equation}
(here 
$ \delta_{ p m } $ 
is the Kronecker symbol,
$ m = 1, 2, \ldots $),
\begin{equation}
\label{eq:int2}
\int\limits_{0}^{b}
r J_1 \left( \frac{ r j_{ 0p } }{ b } \right) H_1^{ ( 0 ) } ( r \tilde{ s } ) dr
=
\frac{ \frac{ 2 i j_{ 0p } }{ \pi b \tilde{ s } } - b \tilde{ s } H_0^{ ( 1 ) }( b \tilde{ s } ) J_1( j_{ 0p } ) }
{ \tilde{ s }^2 - ( j_{ 0p } / b )^2 },
\end{equation}
\begin{equation}
\label{eq:int3}
\int\limits_{0}^{b}
r J_1\left( \frac{ r j_{0p} }{ b } \right) J_1 \left( \frac{ r j_{0m} }{ a } \right) dr
=
\frac{ b \frac{ j_{0m} }{ a } J_0 \left( \frac{ b j_{0m} }{ a } \right) J_1( j_{0p} ) }
{ ( j_{ 0p } / b )^2  - ( j_{ 0m } / a )^2 },
\end{equation}
$ \tilde{ s } = s $
or 
$ \tilde{ s } = s_0 $.
Taking into account that
\begin{equation}
\left( j_{ 0 p } / b \right)^2  - \left( j_{ 0 m } / a \right)^2
=
\left( \gamma_{ z p }^{ ( 1 ) } \right)^2  - \left( \gamma_{ z m }^{(3)} \right)^2,
\end{equation}
\begin{equation}
s_0^2 - \left( j_{ 0 p } / b \right)^2
=
\left( i \omega / V \right)^2 - \left( \gamma_{ z p }^{ ( 1 ) } \right)^2,
\end{equation}
one can obtain after algebraic manipulations:
\begin{widetext}
\begin{align}
\label{eq:sys1}
&\sum\limits_{ m = 1 }^{ \infty }
\left[
\frac{ \tilde{ A }_m }
{ \gamma_{ zm }^{ ( 3 ) } { - } \gamma_{ zp }^{ ( 1 ) } }
{ + }
\frac{ \tilde{ A }_m R_p }
{ \gamma_{ zm }^{ ( 3 ) } { + } \gamma_{ zp }^{ ( 1 ) } }
\right]
{ + }
\frac{ \frac{ i q }{ 2 c b } }{ J_1( j_{ 0 p } ) }
\left[
\left(
\frac{ \pi b^2 s_0^2 h_0 J_1( j_{ 0 p } ) }{ 2 j_{ 0 p } } 
{ - }
1
\right)
\left(
F_{ v p }^{ - } + R_p F_{ v p }^{ + }
\right)
+
F_{ d p }^{ + } + R_p F_{ d p }^{ - }
\right] 
= 0, \\
%
\label{eq:sys2}
&\sum\limits_{ m = 1 }^{ \infty } 
\left[
\frac{ \tilde{ A }_m }
{ \gamma_{ z m }^{ ( 3 ) } { + } \gamma_{ z p }^{ ( 1 ) } }
{ + }
\frac{ \tilde{ A }_m R_p }
{ \gamma_{ z m }^{ ( 3 ) } { - } \gamma_{ z p }^{ ( 1 ) } }
\right]
{ + }
\frac{ \frac{ i q }{ 2 c b } }{ J_1( j_{ 0 p } ) } 
\left[
\left(
\frac{ \pi b^2 s_0^2 h_0 J_1( j_{ 0 p } ) }{ 2 j_{ 0 p } }
{ - }
1
\right)
\left(
R_p F_{ v p }^{ - } { + } F_{ v p }^{ + }
\right)
{ + }
R_p F_{ d p }^{ + } { + } F_{ d p }^{ - }
\right] 
{ = } 
\frac{ 4 \gamma_{ z p }^{ ( 1 ) } \kappa_{ z p }^{ ( 1 ) } \tilde{ B }_p }
{ \kappa_{ z p }^{ ( 1 ) } { + } \varepsilon \gamma_{ z p }^{ ( 1 ) } },
\end{align}
%
\end{widetext}
where
$ h_0 $
is given by Eq.~%
\eqref{eq:h0},
\begin{equation}
\label{eq:tildeA}
\tilde{ A }_m = A_m j_{ 0 m } J_0( b j_{ 0 m } / a ) / a,
\end{equation}
\begin{equation}
\label{eq:tildeB}
\tilde{ B }_p = B_p b J_1( j_{ 0 p } ) / 2 ,
\end{equation}
$ R_p $,
$ F_{ d p }^{ \pm } $
and
$ F_{ v p }^{ \pm } $
are given by Eqs.~%
\eqref{eq:R},
\eqref{eq:Fd}
and
\eqref{eq:Fv},
correspondingly.

In a similar way, we substitute
\eqref{eq:Hphiwide}
and
\eqref{eq:Hphicoax}
into
\eqref{eq:Hphicont2}
and
\eqref{eq:Erocont2},
integrate these relations over the interval
$ b < r < a $ 
with the weight function
$ r Z_p( r \chi_{p} ) $
and utilize the property
\begin{equation}
\label{eq:int5}
\int\nolimits_{b}^{a}
r Z_m( r \chi_{m} ) Z_p( r \chi_p ) dr
=
\delta_{pm}
I_{ p },
\end{equation}
\begin{equation}
\label{eq:Ipp}
I_{ p } 
=
\frac{ a^2 }{ 2 } Z_p^2( a \chi_p ) -
\frac{ b^2 }{ 2 } Z_p^2( b \chi_p ),
\end{equation}
and formulas analogous to
\eqref{eq:int2}
and
\eqref{eq:int3}%
.
Taking into account that
\begin{equation}
\left( j_{0m} / a \right)^2  - \chi_p^2
=
\left( \gamma_{ z m }^{ ( 3 ) } \right)^2  - \left( \gamma_{ z p }^{ ( 2 ) } \right)^2,
\end{equation}
\begin{equation}
s_0^2 - \chi_p^2
=
\left( i \omega / V \right)^2 - \left( \gamma_{ z p }^{ ( 2 ) } \right)^2,
\end{equation}
after a series of algebraic manipulations one obtains
\begin{equation}
\label{eq:sys3}
\sum\limits_{ m = 1 }^{ \infty } \frac{ \tilde{ A }_m }{ \gamma_{zm}^{(3)} - \gamma_{ z n }^{ ( 2 ) } }
+
\frac{ i q }{ 2 c } \frac{ i s_{0}^{2} h_{0} }{ \frac{ \omega }{ i V } - \gamma_{ z n }^{ ( 2 ) } }
= 0,
\end{equation}
\begin{equation}
\label{eq:sys4}
\sum\limits_{ m = 1 }^{ \infty } \frac{ \tilde{ A }_m }{ \gamma_{ z m }^{ ( 3 ) } + \gamma_{ z n }^{ ( 2 ) } }
+
\frac{ i q }{ 2 c } \frac{ i s_{ 0 }^2 h_0 }{ \frac{ \omega }{ i V } + \gamma_{ z n }^{ ( 2 ) } }
= 
- 2 \gamma_{ z p }^{ ( 2 ) } \tilde{ C }_n,
\end{equation}
where
$ n = 0, 1, \ldots $,
\begin{equation}
\label{eq:tildeC}
\tilde{ C }_0 = C_0 \mathrm{ ln }( a / b ),
\quad
\tilde{ C }_p = C_p I_{ p } 
\left[ b Z_p( b \chi_{ p } ) 
\right]^{ -1 }.
\end{equation}
Note that the case 
$ n = 0 $
is obtained by integration of 
\eqref{eq:Hphicont2}
and
\eqref{eq:Erocont2}
over 
$ b < r < a $ 
without any weight function.

In the issue, we obtain four infinite systems
\eqref{eq:sys1},
\eqref{eq:sys2},
\eqref{eq:sys3}
and
\eqref{eq:sys4}.
These systems can be solved simultaneously using the residue-calculus technique~%
\cite{Mittrab,GTV17,GTVA18,GTVGA18}.
This procedure is described in Appendix~%
\ref{App2}.

\section{\label{App2} Constructing the function 
$ f( w ) $
and solving infinite systems.}

\begin{figure}[t]
\centering
\includegraphics[width=0.8\linewidth]{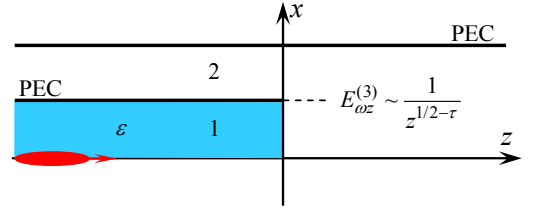}
\caption{\label{fig:Meixner}%
Correct physical behavior of the longitudinal electric field near the sharp edge 
$ r = b $,
$ z \to +0 $
(Meixner edge condition),
$ \tau = \pi^{ -1 } \mathrm{ asin } \frac{ \varepsilon { - } 1  }{ 2 \varepsilon { + } 2 } $.
}
\end{figure}

In accordance with the residue-calculus technique, to solve systems
\eqref{eq:sys1},
\eqref{eq:sys2},
\eqref{eq:sys3}
and
\eqref{eq:sys4}%
, let us  consider the following Cauchy-type integrals over the infinite radius circle
$ C_{ \infty } $:
\begin{equation}
\label{eq:Cauchy}
\oint\nolimits_{ C_{ \infty } } 
\left[
\frac{ f( w ) }{ w { \mp } \gamma_{ z p }^{ ( 1 ) } }
{ + }
\frac{ R_p f( w ) }{ w { \pm } \gamma_{ z p }^{ ( 1 ) } }
\right]
dw
{ = }
\oint\nolimits_{ C_{ \infty } } 
\frac{ f( w ) dw }{ w { \mp } \gamma_{ z n }^{ ( 2 ) } }
{ = }
0,
\end{equation}
where 
$ f( w ) $
is a complex-valued function that should be found.
These integrals equal zero because we suppose that
$ f( w ) $
vanishes for
$ | w | \to \infty $.
Next step is constructing 
$ f( w ) $
so that it has certain specific zeros, poles and behavior for
$ | w | \to \infty $. 
To solve this problem, it is useful to have in mind the infinite systems and their solution for the corresponding vacuum problem (with the same geometry and permittivity
$ \varepsilon = 1 $~%
\cite{GTVA18}%
) and point out the differences.
First, in the case under consideration,  
systems
\eqref{eq:sys1}
and
\eqref{eq:sys2}
are more complicated while systems 
\eqref{eq:sys3}
and
\eqref{eq:sys4}
are the same. 
Second, the singularity of the longitudinal electric field near the sharp edge
$ r = b $,
$ z \to + 0 $,
\begin{equation}
\label{eq:Meixner}
E_{ \omega z }^{ ( 3 ) } \sim 1 / z^{ 1 / 2 - \tau },
\quad
\sin \pi \tau = ( \varepsilon - 1 ) / ( 2 \varepsilon + 2 ),
\end{equation}
becomes weaker in the presence of dielectric
\cite{Mittrab,GTV17}
(see Fig.~%
\ref{fig:Meixner}%
) compared to the vacuum case where we have
$ E_{ \omega z }^{ ( 3 ) } \sim z^{ -1 / 2 } $
near this edge.

%
%

Taking into account these points, one should construct
$ f( w ) $
so that:
\begin{description}
\item[(i)]
$ f( w ) $ 
is regular in complex plane 
$ w $ 
excluding first-order poles 
$ w = \gamma_{ z p }^{ ( 3 ) } $ 
and a pole
$ w = -i \omega / V $;
\item[(ii)]
$ f( w ) $ 
has first-order zeros
$ w = \gamma_{ z n }^{ ( 2 ) } $
and 
$ w = \Gamma_{ m } $,
$ \Gamma_{ m } \ne \gamma_{ z m }^{ ( 1 ) } $%
;
\item[(iii)]
the residue  
$ \mathrm{ Res }_{ -i \omega / V } f( w ) = -q s_0^2 h_0 / ( 2 c ) $;
\item[(iv)]
$ f( w )
{ \xrightarrow[ { | w | { \to } \infty } ] { } } w^{ { - } ( \tau { + } 1/2 ) } $, 
where
$ \sin ( \pi \tau ) { = } \frac{ \varepsilon { - } 1  }{ 2 \varepsilon { + } 2 } $. 
\end{description}
Aforementioned differences from vacuum case are taken into account by items (ii) and (iv).
In the issue, one can write:
\begin{equation}
\label{eq:fapp}
f( w )
=
P
\frac{ ( w {-} \gamma_{ z 0 }^{ ( 2 ) } )
\!\!
\prod\limits_{ s {=} 1 }^{ \infty }
\!\!
\left( 1 {-} \frac{ w }{ \gamma_{ z s }^{ ( 2 ) } } \right)
\prod\limits_{ p {=} 1 }^{ \infty }
\left( 1 {-} \frac{ w }{ \Gamma_{ p } } 
\right) }
{ \left( w - \frac{ \omega }{ i V } \right) 
\prod\nolimits_{ m {=} 1 }^{ \infty }
\left( 1 {-} \frac{ w }{ \gamma_{ z m }^{ ( 3 ) } } \right) } 
Q(w).
\end{equation} 
Here 
$ P $
is unknown constant and 
$ Q( w ) $
is some regular function providing algebraic (instead of exponential) behavior of
$ f( w ) $
for 
$ | w | \to \infty $.
It can be chosen in the same way as in the vacuum case, in accordance with Eq.~%
\eqref{eq:Q}.

Considering integrals
\eqref{eq:Cauchy}
and calculating them using 
\eqref{eq:fapp}
and the residue theorem, we obtain:
\begin{widetext}
%
\begin{align}
\label{eq:Cauchy1}
&\sum\limits_{ m = 1 }^{ \infty }
\left[
\frac{ \mathrm{ Res }_{ \gamma_{ z m }^{ ( 3 ) } } f( w ) }
{ \gamma_{ z m }^{ ( 3 ) } { - } \gamma_{ zp }^{ ( 1 ) } }
{ + }
\frac{ \mathrm{ Res }_{ \gamma_{ z m }^{ ( 3 ) } } f( w ) R_p }
{ \gamma_{ zm }^{ ( 3 ) } { + } \gamma_{ zp }^{ ( 1 ) } }
\right]
{ + }
\left[
\frac{ \mathrm{ Res }_{ \frac{ \omega }{ i V } } f( w ) }
{ \frac{ \omega }{ i V } { - } \gamma_{ zp }^{ ( 1 ) } }
{ + }
\frac{ \mathrm{ Res }_{ \frac{ \omega }{ i V } } f( w ) R_p }
{ \frac{ \omega }{ i V } { + } \gamma_{ zp }^{ ( 1 ) } }
\right]
{ + }
f( \gamma_{ zp }^{ ( 1 ) } )
{ + }
f( - \gamma_{ zp }^{ ( 1 ) } ) R_p
=
0, \\
%
\label{eq:Cauchy2}
&\sum\limits_{ m = 1 }^{ \infty } 
\left[
\frac{ \mathrm{ Res }_{ \gamma_{ z m }^{ ( 3 ) } } f( w ) }
{ \gamma_{ z m }^{ ( 3 ) } { + } \gamma_{ z p }^{ ( 1 ) } }
{ + }
\frac{ \mathrm{ Res }_{ \gamma_{ z m }^{ ( 3 ) } } f( w ) R_p }
{ \gamma_{ z m }^{ ( 3 ) } { - } \gamma_{ z p }^{ ( 1 ) } }
\right]
{ + }
\left[
\frac{ \mathrm{ Res }_{ \frac{ \omega }{ i V } } f( w ) }
{ \frac{ \omega }{ i V } { + } \gamma_{ z p }^{ ( 1 ) } }
{ + }
\frac{ \mathrm{ Res }_{ \frac{ \omega }{ i V } } f( w ) R_p }
{ \frac{ \omega }{ i V } { - } \gamma_{ z p }^{ ( 1 ) } }
\right]
{ + }
f( - \gamma_{ z p }^{ ( 1 ) } )
{ + }
f( \gamma_{ z p }^{ ( 1 ) } ) R_p
= 
0.
\end{align}
%
\end{widetext}
\begin{equation}
\label{eq:Cauchy3}
\sum\limits_{ m = 1 }^{ \infty }
\frac{ \mathrm{ Res }_{ \gamma_{ z m }^{ ( 3 ) } } f( w ) }
{ \gamma_{ z m }^{ ( 3 ) } { - } \gamma_{ z n }^{ ( 2 ) } }
{ + }
\frac{ \mathrm{ Res }_{ \frac{ \omega }{ i V } } f( w ) }
{ \frac{ \omega }{ i V } { - } \gamma_{ z n }^{ ( 2 ) } }
=
0,
\end{equation}
\begin{equation}
\label{eq:Cauchy4}
\sum\limits_{ m = 1 }^{ \infty }
\frac{ \mathrm{ Res }_{ \gamma_{ z m }^{ ( 3 ) } } f( w ) }
{ \gamma_{ z m }^{ ( 3 ) } { + } \gamma_{ z n }^{ ( 2 ) } }
{ + }
\frac{ \mathrm{ Res }_{ \frac{ \omega }{ i V } } f( w ) }
{ \frac{ \omega }{ i V } { + } \gamma_{ z n }^{ ( 2 ) } }
{ + }
f( - \gamma_{ z n }^{ ( 2 ) } )
=
0,
\end{equation}
Note that
\begin{equation}
\label{eq:resomV}
\mathrm{ Res }_{ \frac{ \omega }{ i V } } f( w )
=
P g \left( \frac{ \omega }{ i V } \right),
\end{equation}
where
$ g( w ) $
is given by
\eqref{eq:g}.
Let us compare our systems
\eqref{eq:sys1},
\eqref{eq:sys2},
\eqref{eq:sys3}
and
\eqref{eq:sys4}
with relations
\eqref{eq:Cauchy1},
\eqref{eq:Cauchy2},
\eqref{eq:Cauchy3}
and
\eqref{eq:Cauchy4},
correspondingly.
We put
\begin{equation}
\label{eq:Aapp}
\mathrm{ Res }_{ \gamma_{ z m }^{ ( 3 ) } } f( w ) = \tilde{ A }_m,
\end{equation}
and determine coefficient
$ P $
so that (iii) is fulfilled, i.e.
\begin{equation}
\label{eq:Papp}
P
=
\frac{ i q }{ 2 c }
\frac{ i s_0^2 h_0 }{ g \left( \frac{ \omega }{ i V } \right) }.
\end{equation}
At this step system 
\eqref{eq:sys3}
is formally fulfilled.
Next, we put
\begin{equation}
\label{eq:Capp}
\tilde{ C }_n = f( - \gamma_{ z n }^{ ( 2 ) } )
\left[ 2 \gamma_{ z n }^{ ( 2 ) } \right]^{ -1 },
\end{equation}
\begin{equation}
\begin{aligned}
\label{eq:Bapp}
&- \frac{ i q }{ 2 c b J_1( j_{ 0 p } )} 
\left[
R_p F_{ v p }^{ - } { + } F_{ v p }^{ + }
{ - }
R_p F_{ d p }^{ + } { - } F_{ d p }^{ - }
\right]
- \\
&- \frac{ 4 \gamma_{ z p }^{ ( 1 ) } \kappa_{ z p }^{ ( 1 ) } \tilde{ B }_p }
{ \kappa_{ z p }^{ ( 1 ) } { + } \varepsilon \gamma_{ z p }^{ ( 1 ) } } 
= 
f( - \gamma_{ z p }^{ ( 1 ) } )
{ + }
f( \gamma_{ z p }^{ ( 1 ) } ) R_p,
\end{aligned}
\end{equation}
and systems
\eqref{eq:sys4}
and
\eqref{eq:sys3}
are formally fulfilled as well.
Eq.~%
\eqref{eq:B}
follows from Eqs.~%
\eqref{eq:Bapp}
and
\eqref{eq:tildeB},
Eqs.~%
\eqref{eq:C0}
and
\eqref{eq:C}
follow from
\eqref{eq:Capp}
and
\eqref{eq:tildeC}.
Finally, we put
\begin{equation}
\begin{aligned}
\label{eq:Gammaapp}
- \frac{ i q }{ 2 c b J_1( j_{ 0 p } )} 
&\left[
F_{ v p }^{ - } { + } F_{ v p }^{ + } R_p 
{ - }
F_{ d p }^{ + } { - } F_{ d p }^{ - } R_p 
\right]
= \\ 
&= 
f( \gamma_{ z p }^{ ( 1 ) } )
{ + }
f( - \gamma_{ z p }^{ ( 1 ) } ) R_p,
\end{aligned}
\end{equation}
and the system
\eqref{eq:sys1}
is also fulfilled.
Eq.~%
\eqref{eq:Gammaapp}
is the relation for determination of unknown zeros
$ \Gamma_p $~%
\eqref{eq:Gamma}%
. 
After algebraic transformations it can be rewritten in the following form:
\begin{equation}
\label{eq:infsyst} 
\Delta_{ p } \left(  \left\{ \Delta_{ m }  \right\} \right) = 
\frac{ b }{ \pi }
\frac{ G_{ p } u_{ p } 
\left[ \Gamma_{ p } - \omega / ( i V ) \right]
-
2 \gamma_{ z p }^{ ( 1 ) } R_{ p } }{ \upsilon_{ p + } + R_{ p } \upsilon_{ p - } },
\end{equation}
where
\begin{equation}
\label{eq:up}
u_{ p } \left(  \left\{ \Delta_{ m }  \right\} \right) 
= 
\left.
\frac{ g( w ) } { 1 - \frac{ w }{ \Gamma_{ p } } } 
\right|_{ w = \frac{ \omega }{ i V } },
\end{equation}
\begin{equation}
\label{eq:vp}
\upsilon_{ p \pm } \left(  \left\{ \Delta_{ m }  \right\} \right)
=
\left.
\frac{ f( w ) / P }
{ 1 - \frac{ w }{ \Gamma_{ p } } } 
\right|_{ w = \pm \gamma_{ z p }^{ ( 1 ) } },
\end{equation}
\begin{equation}
\label{eq:Gp}
G_{ p } 
= 
\frac{ F_{ d p }^{ + } + R_{ p } F_{ d p }^{ - } - F_{ v p }^{ - } - R_{ p } F_{ v p }^{ + } }
{ b J_{ 1 }( j_{ 0 p } ) s_0^2 h_0 }. 
\end{equation}
Eq.~%
\eqref{eq:infsyst}
is complicated nonlinear system for
$ \Delta_{ p } $
because expression in the right hand side depends on all unknown
$ \left\{ \Delta_{ m } \right\} $
through
$ u_{ p } $
and
$ \upsilon_{ p \pm } $%
, this fact is underlined by the argument
$ \left\{ \Delta_{ m }  \right\} $
of
$ \Delta_{ p } $,
$ u_{ p } $
and
$ \upsilon_{ p \pm } $. 
This system can be solved numerically using iteration procedure.
Possibility to control the convergence of this procedure is connected with Meixner edge condition
\eqref{eq:Meixner}.

As it was shown in~%
\cite{GTABproc16,GTV17},
condition
\eqref{eq:Meixner}
dictates the following asymptotic behavior of coefficient
$ A_p $
for
$ p \to \infty $:
\begin{equation}
\label{eq:Aas}
A_p \sim p^{ - ( 1 + \tau ) },
\quad
\tilde{ A }_p \sim p^{ - ( 1 / 2 + \tau ) }.
\end{equation}
This in turn results in the asymptotic behavior of 
$ f( w ) $
determined by condition (iv).
Since asymptotic of
$ f( w ) $
is determined by asymptotic of
$ \Gamma_p $, 
$ \gamma_{ z n }^{ ( 2 ) } $
and
$ \gamma_{ z m }^{ ( 3 ) } $ 
for large numbers,
the asymptotic of 
$ \gamma_{ z n }^{ ( 2 ) } $
and
$ \gamma_{ z m }^{ ( 3 ) } $ 
can be easily learned from their definitions
\eqref{eq:gamma2}
and
\eqref{eq:gamma3},
$ \Gamma_p $
should behave as follows for
$ p \to \infty $:
\begin{equation}
\label{eq:Gammaas}
\Gamma_p \sim \frac{ \pi }{ b }
\left( p - 1 / 4 + \tau \right),
\quad
\Delta_p \sim \tau. 
\end{equation}
Therefore, the iteration process for solving
\eqref{eq:infsyst}
is organized as follows. 
We fix quantity 
$ N $ 
of 
$ \Delta_m $,
$ m = 1, 2, \ldots N $ 
to be found. 
For zero-order
approximation, we put 
$ \Delta_m = \tau $ 
for all 
$ m $ 
in the right hand side of
\eqref{eq:infsyst}
and calculate first-order approximation for 
$ \Delta_p $,
$ p = 1, 2, \ldots N $. 
Then we substitute these calculated 
$ \left\{ \Delta_m \right\}$ 
in the right-hand side of 
\eqref{eq:infsyst}
and calculate second-order approximation, etc. 
After these iterations have converged (relative difference in
$ \Delta_N $
for two consequent steps is within the accuracy), we compare 
$ \Delta_N $
with
$ \tau $%
: if 
$ \Delta_N \approx \tau $
within accepted accuracy, process is stopped, otherwise  
$ N $ 
and/or accuracy of calculations is changed and procedure repeats.

\section{\label{App3} Frequency spectrum of the scattered field.}

According to
\eqref{eq:Hphiinner},
\eqref{eq:Hphiwide}
and
\eqref{eq:Hphicoax},
spectrum of the scattered field is determined by spectrum of coefficients
$ \{ A_m \} $, 
$ \{ B_m \} $
and 
$ \{ C_n \} $.
Here we present analytical proving that real spectrum of these coefficients  
contains the same Cherenkov poles as the incident field in the inner dielectric waveguide.
For example, let us consider coefficient
$ \tilde{A}_p $.
In accordance with
\eqref{eq:A}
or 
\eqref{eq:Aapp},
we obtain:
\begin{equation}
\label{eq:tildeAapp}
\tilde{A}_p 
{ = }
P
\frac{ ( \gamma_{ z p }^{ ( 3 ) } {-} \gamma_{ z 0 }^{ ( 2 ) } )
\!\!
\prod\limits_{ n {=} 1 }^{ \infty }
\!\!
\left( 1 {-} \frac{ \gamma_{ z p }^{ ( 3 ) } }{ \gamma_{ z n }^{ ( 2 ) } } \right)
\prod\limits_{ s {=} 1 }^{ \infty }
\left( 1 {-} \frac{ \gamma_{ z p }^{ ( 3 ) } }{ \Gamma_{ s } } 
\right) }
{ \left( \frac{ \frac{ \omega }{ i V } }{ \gamma_{ z p }^{ ( 3 ) } } - 1 \right) 
\prod\limits_{ 
\substack{
		m { = } 1 \\
		m { \ne } p } }^{ \infty }
\left( 1 {-} \frac{ \gamma_{ z p }^{ ( 3 ) } }{ \gamma_{ z m }^{ ( 3 ) } } \right) } 
Q( \gamma_{ z p }^{ ( 3 ) } ). 
\end{equation}
None of the terms in denominator can be zero for real frequencies, therefore only coefficient
$ P $
can have poles.
Definition of 
$ P $
\eqref{eq:P}
or
\eqref{eq:Papp}
can be rewritten as follows:
\begin{equation}
\label{eq:Pappspectrum}
P
=
\frac{ - q s_0^2 h_0 
\prod\limits_{ m { = } 1 }^{ \infty }
\left( 1 {-} \frac{ \frac{ \omega }{ i V } }{ \gamma_{ z m }^{ ( 3 ) } } \right) 
Q( - \frac{ \omega }{ i V } )
}
{ 2 c \left( \frac{ \omega }{ i V } {-} \gamma_{ z 0 }^{ ( 2 ) } \right) 
\prod\limits_{ n {=} 1 }^{ \infty }
\!\!
\left( 1 { - } \frac{ \frac{ \omega }{ i V } }{ \gamma_{ z n }^{ ( 2 ) } } \right)
\prod\limits_{ s {=} 1 }^{ \infty }
\!\!
\left( 1 {-} \frac{ \frac{ \omega }{ i V } }{ \Gamma_{ s } } \right)
}.
\end{equation}
In the denominator, the first term does not equal zero for real 
$ \omega $
and 
$ \beta \ne 1 $,
the first product does not equal zero for real
$ \omega $
as well because
$ \omega / ( i V ) \ne \gamma_{ z n }^{ ( 2 ) } $,
therefore only the second product is a candidate to have real zeros responsible for poles of 
$ P $.
As our numerical results indicate, zeros
$ \Gamma_s $
are specifically shifted in the complex plane so that Eq.~%
\eqref{eq:polesequal}
is fulfilled with high accuracy.
Therefore, coefficient 
$ P $ 
has poles for Cherenkov frequencies
\eqref{eq:Chfreq}.
Since 
$ \{ A_m \} $,
$ \{ B_m \} $
and 
$ \{ C_n \} $
are all proportional to 
$ P $,
the scattered field spectrum contains the same Cherenkov poles 
$ \omega_l^{ \mathrm{ Ch } }$
as the incident field in the area 1, which has to be proved.


%

\end{document}